# Stochastic mechanics, trace dynamics, and differential space – a synthesis

**Mark P. Davidson**
*Spectel Research Corp., 807 Rorke Way, Palo Alto, California 94303*
Email: mdavid@spectelresearch.com

(revised March 21, 2006)

**Abstract**

It is shown how Adler's trace dynamics can be applied to Nelon's stochastic mechanics and also to complex classical dynamical systems. Emergent non-commutivity due to the fractal nature of trajectories in stochastic mechanics is closely related to the fact that the forward and backward time derivatives are different. Consequently, non-commuting matrices representing coordinates and momenta can be defined and then trace dynamics becomes applicable. A new variational approach to stochastic mechanics based on trace dynamics is introduced using this approach. First it is shown that the standard variational methods of stochastic mechanics can be generalized to allow for any diffusion constant in a stochastic model of Schrödinger's equation, and that they can also describe dissipative diffusion. Then it is shown that the trace dynamical theory seems to only describe dissipative diffusion unless an extra quantum mechanical potential term is added to the Hamiltonian. This adds to the debate about what action principle should be applied to stochastic mechanics. It also opens a path for applying trace dynamical method to complex systems which have limiting behavior approximated by Brownian motion. The differential space theory of Wiener and Siegel is reconsidered as a useful tool in this framework, and is generalized to stochastic processes as well as deterministic ones for the hidden trajectories of observables. It also is generalized to a larger class of deterministic models then was previously contemplated. It is proposed that the natural measure space for Wiener-Siegel theory is Haar measure for random unitary matrices. A new physical interpretation of the polychotomic algorithm is given.

PACS Numbers: 03.65.Ta, 05.40.-a, 05.45.-a, 05.45.Df, 02.50-r, 03.65.-w, 03.65.Ta, 04.20.Cv

## I. INTRODUCTION

In a series of papers culminating in a book, Adler and coworkers have introduced and developed a theory which shows how the canonical commutation rules of quantum mechanics can be derived from statistical mechanical arguments applied to dynamics based on traces of matrices[1-5]. Although formally similar to classical mechanics, this theory deals with non-commuting matrices as the basic objects. One of the results is that the emergence of the canonical quantization rules are largely independent of the form of the non-commutivity of the basic matrix elements, as they follow from statistical mechanical arguments which are similar to the equipartition theorem. Adler's motivation, as stated in these works, was to make quantum mechanics more complete. With similar motivation 't Hooft[6-11] has recently proposed that a deterministic and realistic hidden variable model of quantum mechanics could remove much of the mystery surrounding quantum mechanics, avoid the disturbing and (to him) absurd "many worlds" interpretation, and avoid conflict with Bell's theorem and experimental tests of locality. Moreover, he asserts that quantum gravity might be best addressed with this approach; that physics may be deterministic and realistic, and so in essence classical, at the Planck scale; and that in the process of emerging as macroscopic phenomena, information is lost through some collective dissipative, stochastic, or other mechanism resulting in an indeterminate theory which we call quantum mechanics. Recently Penrose too has shown an inclination towards a hidden-variable view of quantum mechanics[12]. All three of these authors dismiss the various no-go theorems regarding hidden variables as essentially irrelevant, citing their similar experience with supersymmetric no-go theorems in particle physics. Adler's and 't Hooft's work is in the tradition of previous efforts in hidden variables including stochastic models[13,14], Bohmian quantum mechanics[15], De Broglie's pilot wave theory[16], Differential Space theory[17-20], etc. The systems they are contemplating are more complex than most of the historical work on hidden variables however, embracing quantum gravity, supersymmetric field theory, and string theory. Other recent work along these lines which deal with stochastic mechanics are the papers by Smolin[21,22] and Santos et al.[23]. A detailed exposition of the interpretive problems of quantum theory and proposed resolutions having stochastic or nonlinear corrections to Schrödinger's equation is found in[24].

There are several different derivations of Nelson's stochastic mechanics from variational principles in the literature[13,25,26,27-29]. An acknowledged deficiency of these methods is the lack of a deeper understanding of the type of classical mechanical system that could justify these action principles. One goal of this paper is to show how Adler's emergent quantum theory is relevant to stochastic mechanics, and to show how a deeper insight into the emergence of stochastic mechanics can possibly be gained from it. Another goal is to show that Adler's theory can be applied to ordinary classical mechanical systems for which the basic objects are commuting, but for which non-commutivity is an emergent property. It is also shown how to generalize the Wiener-Siegel differential space theory to stochastic processes and then how to create stochastic trajectories for quantum mechanical systems as a way to further relate the Adler trace matrix theory to classical mechanics.

For clarity and to avoid excessively complex equations, we shall not consider issues related to Riemannian covariance or magnetic type forces in this paper and we shall write equations for simply one degree of freedom when it is obvious how to extend to many degrees. All of the results here can be extended to the more general cases[30].

## II. ADLER'S TRACE DYNAMICS

We shall only briefly sketch the relevant topics here, and refer the reader to Adler[1] and to references therein. Adler matrices have elements constructed from ordinary complex numbers and from anti-commuting Grassman numbers. In this paper we shall be concerned only with matrices whose elements are ordinary complex-valued functions of time which is a special case. The action principle is expressed in terms of a trace

$$\mathbf{L}[\{q_r\},\{\dot{q}_r\}] = TrL[\{q_r\},\{\dot{q}_r\}] \tag{1}$$

$$\mathbf{S} = \int dt \mathbf{L} \tag{2}$$

where the $q_r$ are self-adjoint matrices and the $\dot{q}_r$ their time derivatives.

Adler defines differentiation with respect to an operator (matrix) in a trace by exploiting the invariance of the trace under cyclic permutations, and so he can make sense of quantities like $\frac{\delta \mathbf{L}}{\delta q_r}$ and $\frac{\delta \mathbf{L}}{\delta \dot{q}_r}$. The principle of least action looks formally similar to the way it does in classical mechanics and one obtains the trace Euler-Lagrange Equations

$$0 = \delta \mathbf{S} = \int dt Tr \sum_r \left( \frac{\delta \mathbf{L}}{\delta q_r} - \frac{d}{dt}\frac{\delta \mathbf{L}}{\delta \dot{q}_r} \right) \delta q_r \tag{3}$$

Defining as usual the conjugate momenta as

$$p_r = \frac{\delta \mathbf{L}}{\delta \dot{q}_r} \tag{4}$$

Then the Euler Lagrange equations take the form

$$\dot{p}_r = \frac{\delta \mathbf{L}}{\delta q_r} \tag{5}$$

And defining a trace Hamiltonian by

$$\mathbf{H} = Tr \sum_r p_r \dot{q}_r - \mathbf{L} \tag{6}$$

It follows that the trace Hamiltonian is a function only of the $p_r$ and $q_r$ and the Hamilton's equations become

$$\frac{\delta \mathbf{H}}{\delta q_r} = -\dot{p}_r, \quad \frac{\delta \mathbf{H}}{\delta p_r} = \dot{q}_r \tag{7}$$

The following matrix or operator is found by Millard[2] to be conserved

$$\tilde{C} = \sum_r (q_r p_r - p_r q_r) = \sum_r [q_r, p_r] \tag{8}$$

In ordinary classical mechanics this would simply vanish. However here we are dealing with non-commuting matrices, and this is a new and non-trivial result for such systems. $\tilde{C}$ is the conserved Noether charge in a matrix model with global unitary invariance and it has vanishing trace

$$Tr\tilde{C} = 0 \qquad (9)$$

as must be true for the trace of any finite sum of commutators in a finite dimensional matrix theory.

A trace dynamics analog of Liouville's theorem is given which shows the invariance of a matrix operator phase space integration under general canonical transformations. The invariant measure is.

$$d\mu = \prod_{r,m,n} d(q_r)^0_{mn} d(p_r)^0_{mn} d(q_r)^1_{mn} d(p_r)^1_{mn} \qquad (10)$$

where the superscipts 0 and 1 denote real and imaginary parts respectively. This measure is invariant even for q's and p's which are not self adjoint matrices. General infinitesimal canonical transformations are defined by letting **G**=TrG, with G self-adjoint but otherwise arbitrary. Then

$$\delta p_r = -\frac{\partial}{\partial q_r}G \quad and \quad \delta q_r = \frac{\partial}{\partial p_r}G \qquad (11)$$

and these leave volume integrals over $d\mu$ and the Poisson brackets in trace dynamics.

It is then assumed that "for a large enough system, the phase space distribution of the matrix variables rapidly loses memory of fine details of their initial values, and that over relevant experimental resolution times the system uniformly samples all phase space configurations that are consistent with the generic conservation laws." In other words the usual ergodic hypothesis is generalized and applied to the present situation. Time averages become ensemble averages. Let $\rho(\{q_r\},\{p_r\})$ be the equilibrium phase space density function. Since $\frac{\partial \rho(\{q_r\},\{p_r\})}{\partial t} = 0$ it must be that $\rho(\{q_r\},\{p_r\})$ depends only on the invariants of the problem. From this the equilibrium limit is described and one can obtain the usual quantum mechanics as a statistical equilibrium. Statistical fluctuations lead to corrections to ordinary quantum mechanics which can be used to test the theory.

### III. STOCHASTIC VARIATIONAL METHODS IN STOCHASTIC MECHANICS

Considering just one dimension for simplicity. The basic generalized Brownian motion process is defined by

$$dx = b(x(t),t)dt + dw(t), \quad E(w(t+dt) - w(t))^2 = 2vdt \qquad (12)$$

It's time reversal also is a Markov process

$$x_*(t) = x(T-t), \quad dx_* = -b_*(x_*(t), T-t)dt + dw_*(t) \tag{13}$$

where $w$ and is a Wiener process, $w_*$ a time reversed Wiener process, b the "forward" drift term, and $b_*$ the "backward" drift term. We use the abbreviated expression $x(t)$ in place of the more correct expression $x(t, \omega)$ where $\omega$ is a point in the underlying probability space as these equations are so common that no confusion likely. Now let us consider Yasue's variational treatment[13,26,28].

Defining an action integral

$$J_{ab} = E\left[\int_a^b \mathsf{L}(x(t), Dx(t), D_*x(t))dt\right] \tag{14}$$

and rendering J stationary under variation of the stochastic process leaving the endpoints at times a and b fixed yields Yasue's version of the Euler-Lagrange equations

$$\frac{\partial \mathsf{L}}{\partial x(t)} - D_*\left(\frac{\partial \mathsf{L}}{\partial Dx(t)}\right) - D\left(\frac{\partial \mathsf{L}}{\partial D_*x(t)}\right) = 0 \tag{15}$$

A simple example of this is a Lagrangian of the form

$$\mathsf{L}(x(t), Dx(t), D_*x(t)) = \frac{1}{2}\left(\frac{1}{2}m|Dx|^2 + \frac{1}{2}m|D_*x|^2\right) - V(x) \tag{16}$$

For which the Euler-Lagrange equations become

$$m\frac{1}{2}(D_*Dx + DD_*x) = -grad\, V(x) \tag{17}$$

This is Nelson's famous version of Newton's law which leads to Schrödinger's equation[14]. Hasegawa[31] has shown that the Yasue Lagrangian is not unique. He considered two cases for the Lagrangian, one leading to dissipative diffusion and the other leading to Schrödinger's equation. He points out that diffusion theory can describe both cases equally well. He shows that the Yasue action also allows for dissipative diffusion which does not yield a Schrödinger equation if the Lagrangian is of the form

$$\mathsf{L}_{Dissipative}(x(t), Dx(t), D_*x(t)) = \frac{1}{2}mDxD_*x - V(x) \tag{18}$$

The central question is what sort of dynamical system would yield a Schrödinger description as opposed to a dissipative diffusion description. Let us consider here another generalization of the Yasue Lagrangian which reduces to the same classical action in the limit of zero diffusion constant

$$\mathbf{L}_G(x(t), Dx(t), D_*x(t)) = \frac{1}{2}m\left(\frac{|Dx|^2 + |D_*x|^2}{2} - \frac{\beta}{8}(Dx - D_*x)^2\right) - V(x) \qquad (19)$$

with $\beta$ a constant. Then the Euler-Lagrange equations become

$$m\left(\frac{(D_*D + DD_*)}{2} + \frac{\beta}{8}(D - D_*)^2\right)x = -\mathrm{grad}\, V(x) \qquad (20)$$

Now if we recall the results of[32] we see that this is equivalent to Schrödinger's equation too provided that

$$v = \frac{1}{\sqrt{1 - \beta/2}}\frac{\hbar}{2m} \qquad (21)$$

And so the diffusion constant is not uniquely defined by the requirement that Schrödinger's equation be derivable from an action principle in stochastic mechanics. This is not an artificial modification of the Lagrangian since it gives exactly the same classical action in the limit of zero diffusion constant. Any value of $\beta$ has as much right as any other to represent the classical Lagrangian as was pointed out in[32]. So Yasue's variational method is compatible with dissipative diffusion and with stochastic models for Schrödinger's equation for any value of the diffusion parameter.

So for any value of the diffusion constant there exists a completely plausible action from which Schrödinger's equation can be derived. This non-uniqueness was known for some time[32-34]. It is known to give Bohm's hidden variable model in the zero diffusion limit[35] and has been explored for large diffusion constants as well to approximate discontinuous processes[36]. It has provided a unifying framework for Bohmian mechanics, stochastic mechanics, and Bell's theory of Beables[37]. It has been extended to time-dependent diffusions as well[38].

An alternative variational method has been given by Guerra and Morato[25]. Following Nelson's treatment[13] of this method the goal is to make some sense of a classical action which is given by

$$I_{classical} = E\int_t^{t_1}\left[\frac{1}{2}\dot{\xi}^2 - \phi\right]ds \qquad (22)$$

where E denotes a probabilistic expectation over all the sample paths. The problem with this is that $\dot{\xi}^2$ is infinite and so defining a suitable action is not straightforward. To proceed, one writes the integral as a particular limiting form. Break the interval [t,t$_1$] into equal segments of duration (t$_1$-t)/N. One defines the action as

$$I_{GM} = \lim_{N \to \infty} E\left[\sum_{a=0}^{N} \left[\frac{1}{2}\left(\frac{\xi(s_a + ds_a) - \xi(s_a)}{ds_a}\right)^2 - \phi\right] ds_a \,\Big|\, \xi(t) = \xi_0\right], \quad (23)$$

$$\text{where } s_a = t + a(t_1 - t)/N \quad \text{and } ds_a = (t_1 - t)/N$$

Then it is shown[13] that

$$E_{x,t} \frac{1}{2}\frac{d\xi(t)d\xi(t)}{dtdt} = \frac{1}{2}b^2 + v\frac{\partial}{\partial x}b(x,t) + \frac{v}{dt} + o(1) \quad (24)$$

$$L_+ = \frac{1}{2}b^2 + v\nabla b - \phi \quad (25)$$

$$I_{GM} = E\int_t^{t_1} L_+(\xi(s), s)ds \quad (26)$$

The infinite term $\dfrac{v}{ds_a}$ does not depend on the variational parameters of the diffusion and so despite the fact that this action is really divergent, a sensible variation can still be carried out. Defining a stochastic analog of Hamilton's principle function by means of the following conditional expectation

$$S_N(x,t) = -E_{x,t}\int_t^{t_1} L_+(\xi(s), s)ds \quad (27)$$

$$DS_N(x,t) = \left(\frac{\partial}{\partial t} + b\nabla + v\Delta\right)S_N(x,t) = L_+(x,t) \quad (28)$$

Varying the diffusion functions leads to the result

$$L_+ - L_+' = b\delta b + v\nabla \delta b \quad (29)$$

and applying the principle of least action one obtains a Hamilton-Jacobi equation for the diffusion

$$\frac{\partial S_N}{\partial t} + \frac{1}{2}(\nabla S_N)^2 + \phi - \frac{v^2}{2}(\nabla \log(\rho))^2 - v^2\Delta \log(\rho) = 0 \quad (30)$$

together with a continuity equation

$$\frac{\partial \rho}{\partial t} + \nabla \cdot \mathbf{v}\rho = 0, \quad \mathbf{v} = \nabla S_N \tag{31}$$

and these are equivalent to Schrödinger's equation provided that $\nu = \frac{\hbar}{2}$ as was shown by Guerra and Morato[13,25].

Nelson has argued for the uniqueness of the diffusion constant based on this fact and other considerations[13]. An effort to generalize this theory was made by Jaekel[27], and he concluded that if one relaxes the Markov condition then a family of stochastic models can be obtained using a slight variation of the Guerra-Morato variational method. These generalizations do allow for an arbitrary diffusion parameter and even a time dependent one. Here we consider another generalization of this method. Consider the action defined by modifying the kinetic energy term

$$I_{MOD} = \lim_{N \to \infty} E \sum_{a=0}^{N} \left[ \frac{1}{2} \left( \frac{\xi(s_a + ds_a) - \xi(s_a)}{ds_a} \right) \left( \frac{\xi(s_a + 2ds_a) - \xi(s_a + ds_a)}{ds_a} \right) - \phi \right] ds_a \, \big| \, \xi(t) = \xi_0, \tag{32}$$

where $s_a = t + a(t_1 - t)/N$ and $ds_a = (t_1 - t)/N$

If the sample trajectories are smooth classical trajectories with continuous first derivatives, then this agrees with the classical action as well as the Guerra-Morato choice does. Consider the kinetic term in the modified action (32). The time intervals in the two velocity factors are no longer overlapping. Therefore the divergent terms will be missing, and we find

$$E_{x,t} \frac{1}{2} \left( \frac{\xi(t+dt) - \xi(t)}{dt} \right) \left( \frac{\xi(t+2dt) - \xi(t+dt)}{dt} \right) = \frac{1}{2} b^2 + o(1) \tag{33}$$

For this modified action we have

$$L_+ = \frac{1}{2} b^2 - \phi \quad \text{and} \quad L_+ - L_+' = b \delta b \tag{34}$$

$$S_{MOD}(x,t) = -E_{x,t} \int_t^{t_1} \left( \frac{1}{2} b^2 - \phi \right) ds \tag{35}$$

$$DS_{MOD}(x,t) = \left( \frac{\partial}{\partial t} + b\nabla + \nu \Delta \right) S_{MOD}(x,t) = \frac{1}{2} b^2 - \phi \tag{36}$$

$$D(S'_{MOD} - S_{MOD}) = D'S'_{MOD} - DS_{MOD} + (D-D')S'_{MOD} = L'_+ - L_+ - \delta b \nabla S_{MOD} + o(\delta b)$$
$$= b\delta b - \delta b \nabla S_{MOD} + o(\delta b) \quad (37)$$

$$I' - I = E\int_t^{t_1} D(S'_{MOD} - S_{MOD})ds = E\int_t^{t_1} (b - \nabla S_{MOD})\delta b + o(\delta b) = o(\delta b) \quad (38)$$

$$b = \nabla S_{MOD} \quad (39)$$

And therefore (36) becomes

$$\left(\frac{\partial S_{MOD}}{\partial t} + \frac{1}{2}\nabla S_{MOD}^2 + \nu\Delta S_{MOD} + \phi\right) = 0 \quad (40)$$

The continuity equation (31) is also satisfied with

$$\mathbf{v} = b - u = \nabla(S_{MOD} - R) = \nabla S_N \quad (41)$$

The Hamilton-Jacobi equation can be written

$$\frac{\partial S_N}{\partial t} + \frac{1}{2}(\nabla S_N)^2 + \phi + \frac{\nu^2}{2}(\nabla \log(\rho))^2 + \nu^2 \Delta \log(\rho) = 0 \quad (42)$$

Compare this result with the original Guerra-Morato result (30). The sign of the quantum mechanical potential term has changed (the two terms involving $\rho$). This gives precisely the dissipative case discussed by Hasegawa[31] and it does not yield Schrödinger's equation. So again, as in the Yasue action approach, we can have an action which gives either Schrödinger's equation or dissipative diffusion. By taking weighted averages of the original Guerra-Morato Lagrangian and the modified one we can construct stochastic models of quantum mechanics with arbitrary diffusion contants as we did in the Yasue case.

## IV. EMERGENT NON-COMMUTING BEHAVIOR IN CLASSICAL SYSTEMS

The position of a classical Brownian particle is simply a c number vector variable, and the positions commute at different times. However there is a temporal non-commutative structure to Brownian motion and it's generalizations[33,34,39]. This structure is closely related to the fact that the sample trajectories of the Wiener process are not differentiable. As a proxy for differentiation, one usually introduces the forward and backward time derivative which are different. It is this difference which is the essential origin of temporal non-commutivity. Moreover this non-commutative structure is closely related to the theory of trace dynamics of Adler, and it is strikingly similar to quantum mechanics. In fact, in stochastic quantum mechanics one can see how the non-commutative structure of quantum mechanics is linked to that of the Brownian motion[34]. It is the non-commutative structure of the diffusion that makes the definition of a unique action ambiguous in both the Yasue sense and the Guerra-Morato sense as we have seen above.

We shall review briefly some of the features of the non-commutative structure of ordinary Brownian motion. Most of this will be taken from[33], but a word of caution. The value of $v$ used there differs by a factor of 2 from the more standard definition used for example in (12). So here we will follow the standard convention and modify the formulas where appropriate. We shall be interested in generalized Brownian motion properties as in (12). Let the Wiener process start at the origin at time 0 so that it has the following properties

$$P(w(t) - w(s) \mid w(r)) = P(w(t) - w(s)), \quad 0 < r < s, t \tag{43}$$

$$E(w(t)) = 0, \text{ all } t \tag{44}$$

$$\rho(w,t) = \frac{1}{\sqrt{4\pi vt}} e^{-\omega^2/4vt}, \text{ where } v \text{ is fixed} \tag{45}$$

$$E(w(t_1)w(t_2)) = 2v \min(t_1, t_2) \tag{46}$$

This last equation is an indication that there is something strange about this process. Taking a derivative with respect to $t_1$ for example we find

$$\frac{\partial}{\partial t_1} E(w(t_1)w(t_2)) = \begin{cases} 2v \text{ if } t_1 < t_2 \\ 0 \text{ if } t_1 > t_2 \end{cases} \tag{47}$$

If we take the limit where $t_1$ approaches $t_2$ from either direction we find two different results. It is a classical result that the time derivative of the Wiener process does not exist[40]. In fact the trajectories of the Wiener process and of stochastic mechanics are known to be Fractal curves[41] with dimension 2. However, the derivatives of expectations do generally exist for this process except in some special cases (for instance when $t_1 = t_2$ above). So for efficient notation let us ignore the fact that strictly-speaking the derivative doesn't exist and write (47) simply as

$$E(\dot{w}(t)w(t \pm \delta)) = \begin{cases} 2v \\ 0 \end{cases}, \delta > 0 \tag{48}$$

The standard way that this eccentric behavior is handled in the mathematics literature is to introduce a forward and backward time derivative which can be defined

equally well for the general Brownian motion process (12). For the Wiener process, these are defined by

$$Dw(t) = \lim_{\Delta t \to 0+} E(\frac{w(t+\Delta t)-w(t)}{\Delta t} | w(t)=w) \qquad (49)$$

$$D_*w(t) = \lim_{\Delta t \to 0+} E(\frac{w(t)-w(t-\Delta t)}{\Delta t} | w(t)=w) \qquad (50)$$

The non-commutative approach, first presented in[33] is another way of looking at this peculiar behavior and it is strikingly similar to quantum mechanics. Defining a time ordered expectation with later times to the right by taking $\delta$ to zero in [79], we have

$$\overrightarrow{\dot{w}(t)w(t)} = 2\nu \qquad (51)$$

$$\overrightarrow{w(t)\dot{w}(t)} = 0 \qquad (52)$$

The arrow gives the direction of increasing time in the taking of limits of the time ordered product. So time increases as we move from left to right if the arrow points this way. This is suggestive of a commutation relation

$$\overrightarrow{[\dot{w}(t), w(t)]} = 2\nu \qquad (53)$$

Let us define a complex Hilbert space $\mathbf{H}_{t_0}$ of functions by giving its inner product as

$$(f,g)_{t_0} = E(f(w(t_0))g(w(t_0))) = \int dw \rho(w,t_0) f(w)g(w) \qquad (54)$$

Note that a particular time is singled out. In the case of a Wiener process the density is a Gaussian and so an orthonormal basis is given by real Hermite Polynomials. Let us define an operator $\hat{\dot{w}}(t_0)$ on $\mathbf{H}_{t_0}$ by

$$(f, \hat{\dot{w}}(t_0)g)_{t_0} = \lim_{\delta_1,\delta_2 \to 0+} E(f(w(t_0-\delta_1))\dot{w}(t_0)g(w(t_0+\delta_2))) \qquad (55)$$

where again we are using our shorthand notation $\dot{w}$ on the right hand side. Let us define an operator $\hat{w}(t_0)$ by simple multiplication. Then it is simple to show[33] that

$$[\hat{\dot{w}}(t_0), \hat{w}(t_0)] = 2\nu \qquad (56)$$

as an operator equation and further that

$$(f, \hat{w}(t_0)g)_{t_0} = \int dw \rho(w,t) f(w) 2\nu \frac{\partial}{\partial w} g(w) \tag{57}$$

So in other words, the velocity operator looks like a derivative acting in configuration space.

$$\hat{w}(t) \Rightarrow 2\nu \frac{\partial}{\partial w} \tag{58}$$

Compare this with quantum mechanics

$$[\hat{q}, \hat{p}] = i\hbar, \; \hat{p} = m\hat{\dot{q}}(t) \Rightarrow -i\hbar \frac{\partial}{\partial q} \tag{59}$$

It's almost the same, except for the "i". For the generalized Brownian motion process, the commutation relations stay the same, but for this case the operator representation becomes[33]

$$\hat{\dot{x}}(t_0) \Rightarrow b(\hat{x}(t_0),t_0) + 2\nu \frac{\partial}{\partial x}, \; \left[\hat{\dot{x}}(t_0), \hat{x}(t_0)\right] = 2\nu \tag{60}$$

Notice that this operator is not self-adjoint.

$$\hat{\dot{x}}(t_0)^\dagger = b(\hat{x}(t_0),t_0) - 2\nu \frac{\partial}{\partial x} \ln(\rho(\hat{x}(t_0),t_0) - 2\nu \frac{\partial}{\partial x} = b_*(\hat{x}(t_0),t_0) - 2\nu \frac{\partial}{\partial x} \tag{61}$$

where $b_*$ is the backward velocity. The operator $\hat{x}(t_0)$ is of course self-adjoint since it acts by just multiplication.

Continuing in this way, an acceleration operator is defined[33] and found to be given by

$$\hat{\ddot{x}}(t_0) = \frac{\partial b(x,t_0)}{\partial t_0} + \nu \frac{\partial^2 b(x,t_0)}{\partial x^2} + \frac{1}{2}\frac{\partial}{\partial x} b^2(x,t_0) \tag{62}$$

This operator commutes with x and since it is a function only of x, it is also self-adjoint. If one defines a stochastic potential by

$$-\frac{\partial \hat{U}}{\partial x} = \hat{\ddot{x}}(t_0) \tag{63}$$

Then the following two equations, called the "Markov Wave Equations" in[33] can be derived for all t

$$\left[2\nu^2 \frac{\partial^2}{\partial x^2} + U\right] e^{R\pm S} = \mp 2\nu \frac{\partial}{\partial t} e^{R\pm S} \qquad (64)$$

where R and S are defined by

$$R = \frac{1}{2}\ln(\rho), \quad b = 2\nu \frac{\partial}{\partial x}(R+S) \qquad (65)$$

These equations were also considered by Hasegawa[31] and by Nagasawa[29]. By defining a Hamiltonian operator by

$$\hat{H} = \frac{1}{2}\hat{\dot{x}}^2(t_0) + \hat{U}(x,t_0) \qquad (66)$$

The expectation of this Hamiltonian is given by

$$(1,\hat{H}1)_{t_0} = \int \rho(x,t_0)\left[\frac{1}{2}\left(b(x(t_0),t_0) + 2\nu\frac{\partial}{\partial x}\right)^2 + \hat{U}(x,t_0)\right]dx \qquad (67)$$

$$(1,\hat{H}1)_{t_0} = \int \rho(x,t_0)\left[\frac{1}{2}\left(b(x(t_0),t_0) - 2\nu\frac{\partial}{\partial x}\ln(\rho(x,t_0))\right)b(x(t_0),t_0) + \hat{U}(x,t_0)\right]dx \qquad (68)$$

but using Nelson notation for the mean and osmotic velocities v and u we have

$$b = u+v, \quad u = \nu \frac{\partial}{\partial x}\ln(\rho) \qquad (69)$$

and so

$$(1,\hat{H}1)_{t_0} = \int \rho(x,t_0)\left[\frac{1}{2}\left(v^2 - u^2\right) + \hat{U}(x,t_0)\right]dx \qquad (70)$$

We see that this is the same formula as Hasegawa's "Dissipative Dynamical Framework" in[31] if. The equations of motion take the form

$$\frac{1}{2}\left(DD + D_*D_*\right)x = -\frac{\partial U}{\partial x} \qquad (71)$$

and if U is taken to be the classical potential then the above does not lead to Schrödinger's equation. However, as was shown in[33] and also[32] if we add an extra term to the potential and modify the Hamiltonian to the following form

$$\hat{H}_{QM} = \frac{1}{2}\hat{\dot{x}}^2(t_0) + \hat{U}(x,t_0) + \alpha \frac{\partial^2 \sqrt{\rho}/\partial x^2}{\sqrt{\rho}}, \quad \alpha \text{ a constant} \tag{72}$$

then the diffusion equations become Schrödinger's equation with a suitable choice for $\alpha$ and for any value of the diffusion constant.

One can calculate the time derivative of any operator by commutation rules with the Hamiltonian

$$\frac{d\hat{f}(\hat{x}(t_0),\hat{\dot{x}}(t_0),t_0)}{dt_0} = \left[\hat{H},\hat{f}\right]/2\nu + \frac{\partial \hat{f}}{\partial t_0} \tag{73}$$

where products of operators are always interpreted as time ordered products at different times in the limit as the times collapse to the same time. Using this together with the commutation rules, we can write a Taylor series for times different from $t_0$ using the recursion

$$\frac{d^N \hat{x}(t_0)}{dt_0^N} = \left[\hat{H}, \frac{d^{N-1}\hat{x}(t_0)}{dt_0^{N-1}}\right]/2\nu + \frac{\partial}{\partial t_0}\left(\frac{d^{N-1}\hat{x}(t_0)}{dt_0^{N-1}}\right) \tag{74}$$

So in this way we can calculate the analog of the Heisenberg operator for all times.

$$\hat{x}(t_0+s) = \sum_{j=0}^{\infty} \frac{d^j \hat{x}(t_0)}{dt_0^j}\frac{s^j}{j!} \tag{75}$$

with a similar expression for the velocity operator at all times and with the assumption that the Taylor series is convergent. An equivalent way of specifying this series is by solving the differential equation

$$\hat{\dot{x}}(t) = \left[\hat{H}(\hat{x}(t),t),\hat{x}(t)\right]/(2\nu) \tag{76}$$

The time evolution operator is a similarity transform which preserves the commutation rules

$$\left[\hat{\dot{x}}(t),\hat{x}(t)\right] = 2\nu \text{ for all } t \tag{77}$$

So now we have been liberated from the time surface $t_0$ and we have defined operators for all times. But notice something surprising. The operators $\hat{x}(t)$ at different times no longer commute. But we started with a stochastic process which was a commuting real diffusion, so how do we interpret this? Consider an expectation of products of operators at different times which in Hilbert space notation given by

$$(1, \prod_{j=1}^{n} \hat{x}(t_j) 1)_{t_0} \qquad (78)$$

Only the time-ordered product makes any sense when comparing to stochastic expectations because we used the time ordering in defining the meaning of the operators. So, we can write a multiple time-ordered Taylor's expansion in the following form

$$E(\prod_{j=1}^{n} x(t_j))_{t_0} = (1, \vec{T} \prod_{j=1}^{n} \hat{x}(t_j) 1)_{t_0} \qquad (79)$$

where the left hand side of this equation is simply the stochastic expectation of a product of commuting random variables, and where $\vec{T}$ is the time ordering operator with time increasing to the right (see Jaekel[39] eqn. 16. This is a generalization of Theorem II and euation 35 in reference[33] for the Wiener process).

We can now define the analog of Heisenberg matrices for any operator O

$$O_{ij} = (\phi_i, \hat{O} \phi_j)_{t_0}, \quad (\phi_i, \phi_j)_{t_0} = \delta_{i,j} \text{ and } \{\phi_i\} \text{ complete} \qquad (80)$$

and we can form traces from these matrices and consider them in the context of Adler's trace dynamics. Note though that the operators $\hat{x}(t)$ and $\hat{\dot{x}}(t)$ are not self-adjoint here. The dynamics of this system are determined by the form of the Hamiltonian. Note also that the trace of a commutator of finite dimensional matrices must vanish, but this seems inconsistent with the commutation rules for p and q. Although finite dimensional matrices representing p and q can never satisfy the canonical commutation rules, they are only approximations to the full operators and are analogous to a truncated Fourier series. The canonical quantization rules can be expected to be approximately safisfied for the lower eigenalue elements of the matrix commutator, but to begin to deviate for the higher ones. By choosing the dimension of the matrices to be larger and larger, the truncation error will go to zero and the commutation relations will be satisfied in the limit of infinite dimension.

## V. STOCHASTIC QUANTUM MECHANICS IN THE CONTEXT OF TRACE DYNAMICS

We shall now use the Adler trace dynamics to derive a new action principle for Stochastic Mechanics. From the previous section we have real matrices representing $x(t)$, $\dot{x}(t)$, and $H(t)$ and so we can calculate the trace of H

$$\mathbf{H} = Tr \hat{H}(\hat{x}, \hat{p}), \quad \hat{p} = \hat{\dot{x}} \qquad (81)$$

We could have chosen to work with the Lagrangian instead, but the results would be the same. In either case there is a problem. Our operators are not self-adjoint. This is true even for the Hamiltonian due to the kinetic energy term. Adler didn't consider non-self-adjoint matrices for the purely c number case. However we also have a real Hilbert space here and so all matrix elements are real. Even for matrices A and B which are not self-adjoint, it is still true that

$$TrAB = TrBA \tag{82}$$

and this cyclic permutivity is all that is required to make sense of operator derivatives, and so Hamilton's equations are still valid and meaningful

$$\frac{\partial \mathbf{H}(\hat{x}, \hat{p})}{\partial \hat{x}} = -\hat{\dot{p}}, \quad \frac{\partial \mathbf{H}(\hat{x}, \hat{p})}{\partial \hat{p}} = \hat{\dot{x}} \tag{83}$$

and in fact, examining the arguments used in trace dynamics, Poisson brackets can still be defined for operators A and B which depend only on $\hat{p}$ and $\hat{x}$ even if they are not self-adjoint

$$\{\hat{A}(\hat{x}, \hat{p}), \hat{B}(\hat{x}, \hat{p})\} = Tr\left(\frac{\delta \hat{A}(\hat{x}, \hat{p})}{\delta x}\frac{\delta \hat{B}(\hat{x}, \hat{p})}{\delta p} - \frac{\delta \hat{B}(\hat{x}, \hat{p})}{\delta x}\frac{\delta \hat{A}(\hat{x}, \hat{p})}{\delta p}\right) \tag{84}$$

It still follows, even for a non-self-adjoint Hamiltonians that

$$\frac{d\mathbf{A}}{dt} = \frac{\partial \mathbf{A}}{\partial t} + \{\mathbf{A}, \mathbf{H}\} \tag{85}$$

The derivation of Liouville's theorem for the measure defined by (generalizing the notation to the case where there are multiple coordinates and momenta)

$$d\mu = \prod_{r,m,n} d(\hat{x}_r)_{mn} d(\hat{p}_r)_{mn} \tag{86}$$

where all the matrix elements are real still is invariant under real canonical transformations.

The commutation rules for stochastic mechanics are already consistent with the emergent commutation relations of trace dynamics because the commutators of each coordinate with its conjugate momenta is a universal constant. The only difference is that the commutators in the stochastic theory are all real, but this is consistent with trace dynamics applied to non self-adjoint matrices.

Now suppose we take the Hamiltonian to be the simplest natural possibility

$$\mathbf{H} = Tr\left[\frac{\hat{p}^2}{2} + U(\hat{x})\right] \tag{87}$$

where $U$ is the classical potential energy. Then applying Hamilton's equations one finds

$$\hat{\dot{x}} = \hat{p}, \quad \hat{\dot{p}} = \hat{\ddot{x}} = -\nabla U(x) \tag{88}$$

and these equations are seen to be identical to the dissipative case found above (71) and in[31,33]. So the Adler trace dynamics seems to lead to dissipative diffusion only and not to Schrödinger's equation unless an extra "quantum mechanical potential" term is added to the Hamiltonian as in (72) and[33]. This puts it at odds with both the Yasue and Guerra-Morato theories which lead naturally to Schrödinger's equation. It is the opinion of the current author that the trace dynamics conclusion is the correct one, and that without the appearance of a new and bizarre potential in the Hamiltonian, the so called "quantum mechanical potential" up to a constant multiplier, quantum mechanics will not evolve statistically from a classical theory. Quantum mechanics does not arise for free simply based on action principle arguments. Although the Yasue and Guerra-Morato approaches are rigorous and self-consistent, they do not derive results from a statistical mechanical argument the way that the Adler trace dynamical theory does. The trace dynamical argument suggests that unless there is an additional force on a particle due to a quantum mechanical potential term, then it would simply settle into a classical Gibbs distribution at some temperature when bound by some attractive potential and it would not in general agree with Schrödinger's equation. The fact that the particle's trajectory is described by the singular and fractal Brownian motion trajectory alone does not alter this conclusion in this author's opinion.

We can have Hermitian operators for coordinates by making an analytic continuation to imaginary values of the diffusion constant[34]. In this way we can recover exactly the mathematics of trace dynamics. First note that the following two equations are equivalent for all nonzero values of complex z [32] (the proof of theorem I was given for real z, but it is trivial to extend it to complex values)

$$\left[-\frac{\hbar^2}{2}\Delta + V\right]\exp(R+iS) = i\hbar\frac{\partial}{\partial t}\exp(R+iS) \tag{89}$$

and

$$\left[-\frac{(z\hbar)^2}{2}\Delta + \left(V + \frac{\hbar^2}{2}(z^2-1)\frac{\Delta\sqrt{\rho}}{\sqrt{\rho}}\right)\right]\exp(R+iS/z) = iz\hbar\frac{\partial}{\partial t}\exp(R+iS/z) \tag{90}$$

Now if we let $z = \pm i|z|$ then (90) becomes

$$\left[\frac{|z|^2\hbar^2}{2}\Delta + \left(V - \frac{\hbar^2}{2}(|z|^2+1)\frac{\Delta\sqrt{\rho}}{\sqrt{\rho}}\right)\right]\exp(R\pm S) = \mp|z|\hbar\frac{\partial}{\partial t}\exp(R\pm S) \tag{91}$$

This is just (64) with $\nu = \frac{\hbar}{2}|z|$ and $U = \left(V - \frac{\hbar^2}{2}(|z|^2 + 1)\frac{\Delta\sqrt{\rho}}{\sqrt{\rho}}\right)$. And wo we can obtain Schrödinger's equation by adding an extra term to the potential of the Markov wave equations for any value of the diffusion constant[33]. See also chapter 4 of Nagasawa in this regard[29]. But we can also obtain it another way. If we analytically continue the diffusion constant to imaginary values but we hold the mean velocity $v\nabla S$ fixed, then (64) becomes the Schrödinger's equation when $\nu = -i\hbar/2$ and when $U = V$. The analytically transformed operators then satisfy the Heisenberg operator algebra (59) and the momentum and position operators become Hermitian. The analytically continued process is a fictitious one that has no physical interpretation. However, it is simple mathematically because the time evolution operator (which is the usual quantum mechanical one in this case) does not depend on the probability density as it would for (91). The Feynman path integral approach is understandable as a mathematical convenience for calculating probabilities in this framework but it too has no physical interpretation. The actual stochastic theory is governed by a nonlinear and non-unique equations, and is much more difficult to do calculations with. Since it is believed that the diffusion constant is not measurable, it is far easier to work with the usual Hermitian operators of quantum mechanics than with the nonlinear Markov theory. In this way a direct link with the Adler Trace dynamics of Hermitian operators can be made.

## VI. DERIVING TRAJECTORIES FROM QUANTUM MECHANICS - WIENER SIEGEL THEORY

It is possible to invert the above procedure and to construct a measurable random trajectory space from quantum mechanics by using the elegant and powerful theory of Wiener and Siegel[18-20,42-44]. We present here a generalization of the Wiener Siegel theory which allows for many random trajectory measures all of which are experimentally indistinguishable from one another, but which provide different stochastic trajectories for each observable of a quantum mechanical system.

The arena of the theory is the same Hilbert space as quantum theory. Besides the usual state vector $\Psi$ which pertains to an ensemble of similarly prepared states, there is another vector in the Hilbert space, call it $\alpha$, which is associated with each element of the ensemble and which is treated like a random variable. Unlike $\Psi$ which is fixed $\alpha$ is a random variable to be averaged over. The probability space for $\alpha$ is called differential space.

The mathematics in the original Wiener-Siegel theory was and is still a hurdle for most physicists. A much simpler theory suffices if the Hilbert space is finite dimensional. This was the starting point of the theory of Bohm and Bub[42,43] who restricted their consideration to finite dimensions and used the "polychotomic" algorithm. We shall only be concerned with finite dimensional matrices here. Wiener and Siegel construct the random $\alpha$ space differently than Bohm and Bub, but both methods are equivalent for finite dimensional matrices. The measure on $\alpha$ space in both cases is constructed so that it is invariant under unitary transformations. So in other words, one can write $\alpha$ as a unitary transformation applied to some fixed vector in the Hilbert space, and then the average can be done with the Haar measure for the unitary group U(N), where N is the dimension of the matrix. So, for an N dimensional Hilbert Space $\mathcal{H}$ we consider a random variable U which maps from a probability space to U(N). Now consider a measurable subset $\mathcal{A}$ of U(N), then

$$\text{Prob}(U \in \mathcal{A}) = \mu_H(\mathcal{A}) \tag{92}$$

where $\mu_H(\mathcal{A})$ is the normalized Haar measure on U(N). Such a variable is called a Haar unitary random variable[45]. The random $\alpha$ is produced by the prescription

$$|\alpha\rangle = U|\alpha_0\rangle \tag{93}$$

with $\alpha_0$ a fixed but arbitrary vector in the Hilbert space. There are two methods desribed in[45] for forming Haar Unitary matrices. One uses Gram-Schmidt orthogonalization on a random Gaussian matrix, the other uses a technique whereby the first column is chosen as a random unit vector, the second column is uniformly chosen from the N-1 dimensional subspace orthogonal to the first column, the third column from the N-2 dimensional subspace orthogonal to the first two columns, etc. The first method is seen to be identical to the Wiener-Siegel Differential space algorithm, and the second is identical to the Bohm-Bub averaging algorithm[44]. Both are clearly equivalent to the Haar measure approach above.

Both Wiener-Siegel and Bohm-Bub assume that the time dependence of the states to be governed by the usual unitary quantum mechanical time evolution operator

$$|\alpha(t)\rangle = U_{QM}(t,t_0)|\alpha(t_0)\rangle \tag{94}$$

$$|\psi(t)\rangle = U_{QM}(t,t_0)|\psi(t_0)\rangle \tag{95}$$

Hidden variables are defined by the polychotomic algorithm. Consider an observable which is represented by a Hermitian matrix A and let $\{\phi_i^A\}$ be an orthonormal basis for the Hilbert space $\mathcal{H}$ formed from the eigenvectors of A.

$$\mathbf{A}|\phi_i^A\rangle = A_i|\phi_i^A\rangle \tag{96}$$

Then the random variable for A is given by the following prescription. Consider the set of ratios

$$R_k(t) = \frac{\left|\left\langle \phi_k^A \middle| \psi(t) \right\rangle\right|}{\left|\left\langle \phi_k^A \middle| \alpha(t) \right\rangle\right|} \tag{97}$$

Find the index k which gives the largest of these ratios (the case where two of the values are exactly the same can be ignored because it is a set of measure zero). The value for the random variable is then the eigenvalue for this kth eigenvector.

$$A(\alpha(t), \psi(t)) = A_{k(t)} \tag{98}$$

As time evolves, k(t) will jump around among the possible eigenvalues of the operator **A**, and thus map out a trajectory for the observable A. This polychotomic algorithm maps the Haar unitaries to real numbers, and the probability density is the same as quantum mechanics as was shown by Wiener and Siegel and also by Bohm and Bub

$$\text{Pr}\,ob(k(t) = \tilde{k}) = \left|\left\langle \phi_{\tilde{k}}^A \middle| \psi(t) \right\rangle\right|^2 \tag{99}$$

So we can create a stochastic process for any observable in this way.

The time evolution conditions are arbitrary. We can clearly modify the time evolution of $|\alpha(t)\rangle$ to the following without affecting this probability result

$$|\alpha(t)\rangle = U_R(t, t_0)|\alpha(t_0)\rangle \tag{100}$$

where $U_R$ is any unitary matrix function of t and $t_0$. This is because the Haar measure is invariant under unitary transformations and this modification will therefore not affect any averages over the initial microstate $\alpha(t_0)$. All of the probabilities derivable from the polychotomic algorithm will remain unchanged no matter what choice of $U_R(t, t_0)$ is made. It is universally accepted that if the probability distributions for all observables are not affected by a transformation, then the transformation is not a measurable. Examples are gauge transformations or the overall phase of a quantum state vector. Here we have a much higher order of unobservable complexity, since the number of possible functions $U_R(t, t_0)$ are truly enormous. The sample trajectories for observables will depend on the choice of $U_R(t, t_0)$.

The traditional Wiener Siegel theory is a deterministic theory. However, by exploiting the above non-uniqueness, we can convert it into a stochastic theory by letting $U_R$ become any random unitary stochastic matrix. If the $U_R$ are a time dependent Markov process then the trajectories for all observables will also be Markov processes. It is likely that for some choice of $U_R$ the trajectories for a single particle will resemble stochastic mechanics. But there has been very little work done on trajectory characterization for the polychotomic algorithm and Wiener-Siegel theory. The subject of random unitary matrices relates to number theory, the Riemann zeta function, and other areas of physics, and so there is a large body of relevant mathematical literature on it[46].

One problem with the polychotomic algorithm is how to give it a physical interpretation. Here then is a possible one. Since $|\alpha(t)\rangle$ and $|\psi(t)\rangle$ are both elements of the same Hilbert space, it is possible to add them together. This suggests, since $|\alpha(t)\rangle$ is a random variable, that it is a noise term to be added to the "signal" $|\psi(t)\rangle$. So we can think that each member of an ensemble of similarly prepared states has a state vector given by

$$|\Psi_p\rangle = \frac{|\psi(t)\rangle + \lambda |\alpha_p(t)\rangle}{N} \tag{101}$$

where $\lambda$ is a small constant multiplier and N is a normalization constant. Then the polychotomic algorithm looks like it is maximizing the signal to noise ratio between the projection of these two terms onto a set of basis vectors which are eigenvector of an observable in question.

One of the general predictions of Adler's trace dynamics is the existence of stochastic noise on the Schrödinger equation. Conventional quantum mechanics is seen as a type of thermodynamic equilibrium about which there are fluctuations. And so the interpretation of the polychotomic algorithm of the Wiener-Siegel theory as a maximization of a signal to noise term for the slightly fluctuating state of a system seems at least qualitatively very consistent with the Trace dynamics theory.

Note that the Wiener-Siegel theory and its generalization works equally well for quantum systems with anticommuting operators and Fermions. However, it yields trajectories only for observables, that is for Hermitian operators.

So classical physics can result in chaotic complex behavior which generates non-commutative temporal behavior and then Adler's theory shows us how to treat such systems statistically, resulting in commutation rules similar to quantum mechanics, and then finally the Wiener Siegel theory gives us a way to go back and reconstruct trajectory models for the theory we started out with. So we can go either way to and from the different descriptions, somewhat like complementarity in quantum mechanics. Perhaps both Bohr and Einstein were right in their debates on quantum mechanics. Perhaps quantum mechanics is indeed incomplete, as Einstein asserted, but there are many experimentally equivalent "hidden variable" descriptions allowed too, as one might imagine Bohr to have asserted. Each choice for the random unitary $U_R$ gives a possible different hidden variable model which is equivalent to quantum mechanics, just as different diffusion parameters can be used to model quantum mechanics in stochastic mechanics. So there is a high degree of non-uniqueness in portraying quantum mechanics as a hidden variable theory whether it be deterministic or stochastic. This impasse awaits an experimental failure of quantum mechanics which will enable us to distinguish them and tell the right one.

One advantage that the Wiener-Siegel theory has over stochastic mechanics is that it provides a stochastic process for every single observable operator, whereas stochastic mechanics singles out the coordinate variable.

## VII. CONCLUSION

Trace dynamics, stochastic mechanics, and differential space theory are useful in combination for exploring the stochastic nature of quantum mechanics as well as the behavior of complex classical systems.

Applying trace dynamics to a classical system which has emergent non-commutivity leads most naturally to a system with dissipative dynamics in the sense of Hasegawa, and not to quantum mechanics. So one must add an unusual term to the potential energy which is proportional to Bohm's quantum mechanical potential to get Schrödinger's equation out. Now we have three variational derivations of the equations of motion for stochastic mechanics. Guerra and Morato's lead naturally to Schrödinger's equation with a unique value for the diffusion parameter, but it can be modified to yield dissipative behavior and mixed Schrödinger models with arbitrary diffusion parameter. Yasue's variational method can yield a model for Schrödinger's equation for any diffusion parameter, or it can lead to dissipative diffusion. And finally Adler's theory applied to stochastic mechanics would seem to yield only a dissipative dynamical system unless an extra term is added to the potential energy. The author favors this last conclusion. He believes that ordinary classical systems will not yield quantum mechanical behavior without some unusual force which leads to the presence of the extra potential which is Bohm's quantum mechanical potential up to a constant. Examples which might yield an extra force term of the right form are usually quite bizarre. For example, a radiation reaction force model leads to it[47] but also to an extra nonlinear term[48], and perhaps a bound state tachyon model might lead to such behavior [49].

A question to ponder is what classical mechanical systems might possess emergent non-commutivity in the sense used here and therefore be a candidate for trace dynamics. Perhaps a turbulent classical fluid in the limit of fully developed isotropic and homogeneous turbulence might be a candidate. Test particles can be immersed in such a fluid and their motion tracked electronically, or such systems can be simulated numerically. Naturally the Wiener process could be only an approximation to the small time behavior, and so the time discreteness constant would have to be suitably chosen. In the infinite stirring rate limit the time constant might approach zero however. By studying the two particle dispersion[50] the small distance structure of the diffusion induced by turbulence can be examined. We propose that this system will show both emergent non-commutative behavior as well as the dissipative statistical mechanical behavior predicted by trace dynamics. This could be a laboratory for exploring trace dynamical theory. Another simple system that might be applicable is simply a Newtonian billiard ball model for a gas. This simple kinetic theory ought to exhibit Markovian diffusion behavior which should allow for emergent non-commutivity. Numerical experiments could be used to explore trace dynamical arguments.

Although we have shown how trace dynamics seems to favor dissipative diffusion in the current context, it's still possible that some other way of looking at the operator space might more naturally allow the theory to include Schrödinger's equation as an option built into the variational principle as was done for the other variational methods. The fact that the operators and their adjoints are different allows one to construct self-adjoint operators by taking sums and differences and adding them together with complex coefficients. Perhaps such an approach could allow the introduction of self-adjoint operators into the theory in a natural way and thus extend the action principle of trace-dynamics to quantum mechanics without needing to add an extra term.

An open question is how to include the Fermionic sector of Adler's matrix theory as emergent from a classical commuting theory as has been done here for the Bosonic sector. There has been some work on extending diffusion theory to anticommuting operator spaces[51-53], and perhaps this could be a basis for extending the ideas presented here to include fermions. But it would mean that Fermions do not have the same visualizable trajectories that the purely Boson diffusions have.

**Acknowledgements**

The author acknowledges helpful correspondence with Basil Hiley on the Bohm-Bub theory.